\documentclass[12pt,aps,noshowpacs]{revtex4}
\usepackage{epsfig,amssymb,amsfonts,bm,psfrag}

\usepackage[active]{srcltx} 

\newcommand{\vp}{\varphi}
\newcommand{\be}{\begin{equation}}
\newcommand{\ee}{\end{equation}}
\newcommand{\bea}{\begin{eqnarray}}
\newcommand{\eea}{\end{eqnarray}}

\begin{document}

\title{QCD relics from the early Universe}

\author{D. Antonov}
\affiliation{Departamento de F\'isica and Centro de F\'isica das Interac\c{c}\~oes Fundamentais,
Instituto Superior T\'ecnico, UT Lisboa, Av. Rovisco Pais, 1049-001 Lisboa, Portugal}

\author{A.V. Nefediev}
\affiliation{Institute of Theoretical and Experimental Physics, B. Cheremushkinskaya 25, 117218 Moscow, Russian
Federation}

\author{J.E.F.T. Ribeiro}
\affiliation{Departamento de F\'isica and Centro de F\'isica das Interac\c{c}\~oes Fundamentais,
Instituto Superior T\'ecnico, UT Lisboa, Av. Rovisco Pais, 1049-001 Lisboa, Portugal}

\begin{abstract}
We suggest the possibility of creation in the early Universe of stable domains of radius a few
kilometers wide, formed by coherently excited states of $\pi$-mesons. Such domains appear dark to an external observer, since the decay rate of the said coherent pionic states into photons is vanishingly small. The related thermal insulation
of the domains from the outer world could have allowed them to survive till present days. The estimated maximum radius and the period of rotation of such objects turn out to be compatible with those of certain pulsars.
\end{abstract}

\maketitle


\section{Introduction}

Low-energy QCD can be characterized by four nonperturbative quantities, of which the gluon condensate
$\left<g^2(F_{\mu\nu}^a)^2\right>\equiv\left<G^2\right>$ and the vacuum correlation length $a$ are 
related to confinement (so that the string tension between the two static quarks in the fundamental 
representation is $\sigma_{\rm f}\propto a^2\cdot\left<G^2\right>$~\cite{ds}), while the chiral 
condensate $\left<\bar\psi\psi\right>$ and the constituent quark mass $m$ characterize spontaneous 
breaking of chiral symmetry. One can still consider an extrapolation of the chiral condensate 
$\left<\bar\psi\psi\right>$ to the heavy-quark limit (where $\left<\bar\psi\psi\right>_{\rm heavy}$ 
ceases to be an order parameter of spontaneous chiral symmetry breaking). In this limit, 
the role of the constituent mass of the quark is played by its current mass $M$, 
yielding the known formula of QCD sum rules~\cite{svz}: 
$\left<\bar\psi\psi\right>_{\rm heavy}\propto -\left<G^2\right>/M$. 
In fact, as long as the mean size of the quark trajectory in Euclidean space remains smaller than $a$, the nonperturbative 
Yang--Mills fields inside the trajectory can be treated as constant, leading to the area-squared law 
for the quark's Wilson loop~\cite{ds}.
As has been shown in Ref.~\cite{AntRib}, this law reproduces correctly the above-mentioned 
heavy-quark condensate $\left<\bar\psi\psi\right>_{\rm heavy}$. With the decrease of the 
current quark mass, the mean size of the quark trajectory gradually gets larger than $a$, and the 
area-squared law of the Wilson loop gets morphed into an area law. The effective scale-dependent 
string tension entering the area law characterizes the strength of the confining force between the 
light quarks much in the same way as the constant string tension does for heavy quarks. 

Alternatively, one can use Nambu--Jona-Lasinio-type models with confining kernels, henceforth called Generalized NJL (GNJL) models, as 
a tool to study the interplay between confinement and spontaneous chiral symmetry breaking. 
Unlike the standard Nambu--Jona-Lasinio (NJL) models~\cite{NJL} with either local or short-ranged kernels, 
which do not describe confinement, the latter is present explicitly in GNJL models~\cite{results, fr, p}, which successfully fulfill
the known low-energy theorems of Gell-Mann, Oakes and Renner~\cite{GMOR}, Goldberger and 
Treiman~\cite{GT}, Weinberg~\cite{Weinberg}, and reproduce the so-called Adler zero~\cite{Adler}.
Yet another positive property of such models is the stability of the said results against 
variations of the parameters, as well as the functional form of the confining kernel. Therefore we are led 
to conclude that these models do provide a robust mechanism for the existence of a Goldstone boson --- the
pion --- which becomes massless in the chiral limit. The presence of this boson can be associated with 
the specific realization of spontaneous chiral symmetry breaking, which is related, through the confining kernels, to confinement. This mechanism is enshrined in what is known as the  mass-gap equation (MGE).
Solving this equation amounts to a proper choice of the quark Fock space $\cal {F}$, that is, to a 
proper choice of the fermionic ground state of the theory. In GNJL models, the MGE
turns out to possess more than one solution, the replicas~\cite{fr}. Among those, the one with the 
lowest energy yields the QCD chiral condensate. Higher replicas describe excited states, in which the 
absolute value of the chiral condensate is smaller (that is, the chiral symmetry is ``less broken'') than in the 
QCD vacuum. Therefore, one can naturally pose a question as how and where such excited states 
could exist in Nature.  Because theoretical evidence for the possible existence of such states (which, as we shall see,
can be realized as coherent states of pions) was restricted to 
GNJL calculations, it is important to verify, within the effective-action approach, which is a gauge-invariant formalism, 
whether such states may remain a degree of freedom of QCD. To this end, we study in Sec.~II the
correspondence between the effective-action approach of Ref.~\cite{AntRib} and GNJL models. 
In this paper, we show that, despite their different nature, both approaches turn out to possess, in the chiral limit, 
a scale-dependent effective mass, which is necessary condition for the existence of replicas. 
In the heavy-quark limit, the two approaches have already been shown equivalent, cf. Ref.~\cite{AntRib}.

As for the actual problem of possible realization of replicas in Nature, we argue 
that such replicas, once thermally excited in the 
early Universe, may form certain stable macroscopic domains. 
To explore such domains quantitatively, we calculate in Sec.~III their internal energy. 
It is given by the difference between the internal energies of the 
replica-filled domain and the physical vacuum outside that domain. As such, 
this internal energy is positive-definite, making the domains stable against the gravitational collapse.
Owing to this fact, the domains can be studied by using the well-known
Tolman--Oppenheimer--Volkoff equation~\cite{w}. For sufficiently 
low temperatures, we show the pionic contributions to the replicas' internal energy to be 
negligible when compared with the ``excited vacuum'' contributions, thence allowing us to assume the internal 
energy inside such domains to be constant.

This assumption allows us to apply the known form for the metric inside spherically-symmetric objects with 
a constant energy density (as provided by the corresponding solution to the 
Tolman--Oppenheimer--Volkoff equation~\cite{w}) to calculate quantitatively the leading gravitational 
correction to the chiral condensate inside the domains. As a result, we find that this correction could only 
become relevant 
at the distances $\lesssim 0.55{\,}{\rm fm}$ to the center of the replica-filled domain, which are, for the first 
excited replica, 20 orders of magnitude 
smaller than the size of the domain. Such irrelevance of the gravitational correction to the chiral condensate 
makes our approach selfconsistent, since it justifies 
our use of the flat-space internal energy of the first replica.
Furthermore, by a simple estimate we show that the anomalous decay-rate of coherently excited 
$\pi$-mesons, which fill the domain, into two photons is vanishingly small. Using this fact, we put forward a 
conjecture that such domains in the Universe have had a chance to survive till the present time. 

The paper is organized as follows. Section~II is devoted to a theoretical discussion of replicas. We show how 
a long-range confining kernel can be modeled, in both GNJL and the effective-action approaches, by a scale-dependent 
constituent quark mass, which, in turn, becomes the cornerstone for the possible existence of replicas.
In Sec.~III, we calculate the internal energy of the domain filled with coherently excited 
$\pi$-mesons, as well as its maximum radius compatible with the stability against the gravitational collapse, 
and a possible period of rotation. In Sec.~IV, we use the effective-action approach to evaluate gravitational 
corrections to the chiral condensate, and show such corrections to be negligible. 
In Sec.~V, we summarize the main results of the paper.
In Appendix~A, we recall the world-line representation of the Abelian Higgs model with a nonlocal 
quartic Higgs interaction, that is the bosonic prototype of an NJL model with a nonlocal kernel.
In Appendix~B, for the sake of completeness, we collect well-known results on the gravitational field of a 
spherically-symmetric object with a constant energy density.

\section{Replicas}

The possible existence, in QCD, of new types of physical states, hitherto named replicas, has been put forward by phenomenological studies based on the GNJL models~\cite{results, fr, MoreRep, rep, replica2d}.  
In this paper, we follow two pronged approaches to the issue of replicas. First, in Subsec.~\ref{IIA} below, we give a bird's eye 
view on the physics and the origin of replicas within GNJL models. Then, in Subsec.~\ref{IIB}, we proceed to establish the 
connection between the GNJL and the effective-action approaches. This is important not only because of the relevance of 
validating, within a different setting, the possible existence of replicas, but also because, once such states are 
to fill a macroscopic domain (and hence to be a source of gravity), it is 
of paramount importance to study the back-reaction, which the gravity might have on the quark condensate (that has produced, 
in first instance, such gravity forces). It happens that this study can reliably be carried out within the 
effective-action formalism -- see Sec.~IV -- which is not the case for GNJL models. 

\subsection{The microscopic analysis of spontaneous chiral symmetry breaking and the origin of replicas}\label{IIA}

In order to illustrate the correspondence between the GNJL and the effective-action approaches to chiral symmetry breaking, 
let us start with a brief introduction to GNJL models of QCD. Upon integration in the QCD partition function over 
the gluonic fields $A_\mu^a$ in the Gaussian approximation, that is, disregarding the explicit 
contributions of all connected averages  
$\bigl<g^n A_{\mu_1}^{a_1} A_{\mu_2}^{a_2}\cdots A_{\mu_n}^{a_n}\bigr>_{\rm conn}$ with $n>2$, one obtains 
an effective action containing a four-quark interaction:
\begin{equation}
\label{njl0}
S={\cal T}\left[ \int_x \bar\psi(\gamma_\mu\partial_\mu+m)\psi+\frac12\int_{x,y} j_\mu^a(x){\,}
\bigl<g^2A_\mu^a(x)A_\nu^b(y)\bigr>_{\rm conn}{\,}j_\nu^b(y)\right].
\end{equation}
Here ${\cal T}$ stands for time ordering, $j_\mu^a\equiv\bar\psi\gamma_\mu T^a\psi$ is the quark current, 
$T^a$ is an SU($N_c$)-generator in the fundamental representation, and we use a shorthand notation $\int_x\equiv\int d^4x$. 
Of course, the heavier the current quark masses $m$ are, the better the Gaussian 
approximation becomes, in which case the two-point connected correlation function 
$\bigl<g^2A_\mu^a(x)A_\nu^b(y)\bigr>_{\rm conn}$ becomes the usual gluon propagator 
$\bigl<g^2A_\mu^a(x)A_\nu^b(y)\bigr>$. For lower current quark masses, $\bigl<g^2 A_\mu^a(x)A_\nu^b(y)\bigr>$ must be 
thought as representing an effective, phenomenologically given, force between quarks. Let us denote
\be\label{massdep}
D_{\mu\nu}^{ab}(x,y)=\bigl<g^2 A_\mu^a(x)A_\nu^b(y)\bigr>,~~~  
\tilde{m}(x)=m+2\int_y D_{\mu\nu}^{ab}(x,y)\gamma_\mu T^a{\,}\bigl< 
\psi (y)\bar\psi (x)\bigr>_{\rm w}{\,}\gamma_\nu T^b,
\ee 
where $\bigl<\ldots\bigr>_{\rm w}$ stands for the Wick contraction.
Using the Wick theorem, we can then cast Eq.~(\ref{njl0}) to the form
\be\label{njl2}
S=S_0 + {\,}\colon \int_x\bar\psi\left[ \gamma_\mu\partial_\mu+\tilde{m}(x) \right] \psi+
\frac12\int_{x,y} j_\mu^a(x)D_{\mu\nu}^{ab}(x,y)j_\nu^b(y) \colon{\,},
\ee
where $S_0$ is a $c$-number resulting from the Wick contractions, 
and $\colon$~$\colon$ stands for normal ordering. There exists a vast literature devoted to 
phenomenological ans\"atze for  
$D_{\mu\nu}^{ab}(x,y)$~\cite{results, fr}, aimed at the simulation of confining forces between quarks 
(and hence at the simulation of the contributions of the entirety of the cumulant expansion to the quark confining 
forces). 
It turns out that, regardless of the form of $D_{\mu\nu}^{ab}(x,y)$, 
we have the effective quark mass $\tilde{m}(x)\simeq M$ for the heavy-quark current mass $M$.  However, 
when the current quark mass $m\rightarrow 0$, $\tilde{m}(x)$  does not in 
general go to zero but rather becomes \emph{a function of the quark-antiquark separation}.
GNJL models provide one with a natural tool for studies of this phenomenon. Indeed, considering an instantaneous correlation function $D_{\mu\nu}^{ab}(x,y)$ given by a suitable confining potential, one can apply to the problem the whole variety of Hamiltonian-based methods known in the literature. 

The second-quantized Hamiltonian corresponding to action (\ref{njl0}) contains 
anomalous (also called off-diagonal) terms. Among those, the bilinear anomalous terms coming 
from $\colon \int_x\bar\psi\left[ \gamma_\mu\partial_\mu+\tilde{m}(x) \right] \psi\colon $ are of particular importance. 
In order to eliminate such terms, one  uses a  
Bogoliubov--Valatin transformation ${\cal VB} [\vp_{\bf p}]$  \cite{fr} to represent the original quark fields 
via new ``dressed`` quark fields. The transformation ${\cal VB} [\vp_{\bf p}]$ can be 
thought of as a rotation in the Hilbert space spanned by the $u$ and $v$ spinors (which together 
compose a bispinor $\psi$) parametrized by a momentum-dependent angle $\vp_{\bf p}$. 
Within GNJL models, one further eliminates the anomalous terms by means of the 
mass-gap equation ($MGE[\vp_{\bf p}]=0$), whose solution(s) define the chiral angle $\vp_{\bf p}$. 
It is therefore quite clear that \emph{the prerequisite for having non-trivial solutions for $\vp_{\bf p}$ is the 
existence of a non-zero $\tilde{m}(x)$}.  The remaining quartic anomalous terms can also be eliminated by another 
Bogoliubov--Valatin transformation --- see the discussion at the end of this subsection.

Now, the mass-gap equation, being nonlinear, may have multiple
multi-node solutions for the chiral angle $\vp_{\bf p}$ -- the replicas. 
For the quadratic kernel, the replicas 
were discussed already in the pioneering papers~\cite{fr}, while the long-range kernels of a generic form and 
2D QCD (the 't~Hooft model) in this context were studied
in Refs.~\cite{MoreRep, rep} and Ref.~\cite{replica2d}, respectively.
Existence of multiple solutions to the MGE was also confirmed by the calculations done
in other approaches~\cite{OH}. The zeroth replica  
$\vp^{(R_0)}_{\bf p}$ provides the vacuum state $|0\rangle$ with the lowest energy-density, and defines the 
corresponding fermionic Fock space ${\cal F}_0$. The vacuum energy-density corresponding to 
any other Fock space ${\cal F}_n$ is larger. For ${\cal F}_0$, the MGE 
coincides with the pion Salpeter equation and ensures, at the same time, both the chiral pion 
to be massless and $\pi -\pi$ scattering lengths to be zero. In other words, \emph {it is 
in general unavoidable to have a chiral 
pion plus all the low-energy theorems thereto, and not to have an $n$-numerable set of 
solutions $\left\{\vp^{(R_n)}_{\bf p},\;\bigl<\bar\psi\psi\bigr>_n\right\}$ to the MGE}.

An operator creating a certain replica $R$ can be written in a closed form as~\cite{rep} 
\be
|R\rangle=S|0\rangle,\quad
S=\mathop{\prod}\limits_{c,f,{\bf p}}\left(\cos^2\frac{\Delta\vp_{\bf p}}{2}+
\sin\frac{\Delta\vp_{\bf p}}{2}\cos\frac{\Delta\vp_{\bf p}}{2}C^\dagger_{c,f,{\bf p}}
+\frac12\sin^2\frac{\Delta\vp_{\bf p}}{2}C^{\dagger 2}_{c,f,{\bf p}}\right),
\label{nv2} 
\ee 
where $|0\rangle$ and $|R\rangle$ are the
physical BCS vacuum and the vacuum of the replica, which are defined by the chiral angles
$\vp^{(0)}_{\bf p}$ and $\vp^{(R)}_{\bf p}$, respectively, and 
$\Delta\vp_{\bf p}=\vp^{(R)}_{\bf p}-\vp^{(0)}_{\bf p}$. The operator $C_{c,f,{\bf p}}^\dagger$
creates a $^3P_0$ quark-antiquark pair of color $c$, flavor $f$,
and with the relative 3-momentum ${\bf p}$: 
\be
C_{c,f,{\bf p}}^\dagger=[b^\dagger_{\uparrow
cf}({\bf p}),b^\dagger_{\downarrow cf}({\bf p})]{\,}
{\cal M}_{^3P_0}
\left[\begin{array}{c}d^\dagger_{\uparrow cf}({\bf p})\\
d^\dagger_{\downarrow cf}({\bf p})\end{array}\right],
\quad{\cal M}_{^3P_0}=(\vec{\sigma}\hat{{\bf p}})i\sigma_2,
\label{Cddef} 
\ee 
where $\sigma$'s are the Pauli
matrices.
It turns out that, for a sufficiently large spatial volume $V^{(3)}$, the states $|0\rangle$ and $|R\rangle$
are essentially orthogonal to each other~\cite{rep}. That is because their overlap gets exponentially suppressed with the increase of the volume $V^{(3)}$, namely
\be
\langle 0|R\rangle=\exp\left[N_cN_{\rm f}V^{(3)}\int\frac{d^3p}{(2\pi)^3}
\ln\left(\cos^2\frac{\Delta\vp_{\bf p}}{2}\right)\right]. 
\label{norm}
\ee 
The same is obviously true for the overlap of any two quark states belonging to different Fock spaces.

Furthermore, as was shown in Ref.~\cite{MoreRep}, for any
chiral angle respecting the MGE, yet another 
Bogoliubov--Valatin-like transformation can be performed in order 
to reformulate the GNJL model in terms of the 
observable mesonic states. Such a transformation can be applied 
to replicas as well, provided the volume $V^{(3)}$ is large enough. 
As a result, the quark BCS states $|q\rangle_0$ built on top of the vacuum $|0\rangle$ get transformed to the mesonic states $|M\rangle_0$ built on top of the mesonic vacuum  
$|\Omega\rangle_0$, while the replica quark states $|q\rangle_R$ get transformed to some coherent 
mesonic states $|M\rangle_R$ --- excitations over $|\Omega\rangle_R$. It turns out that the operator $S$ defined in  Eq.~(\ref{nv2})
can be represented in terms of the compound mesonic operators. 
Since the dominant role in the cloud of coherent mesonic excitations 
is played by the lightest hadrons, 
it is natural to approximately consider mesonic replica-states as being formed by 
``almost non-interacting" pions. The weakness of pionic interactions 
is a consequence of chiral symmetry. Namely, as was already mentioned, for massless quarks, the $\pi-\pi$ scattering 
length vanishes, and the pions at low momenta appear almost transparent to each other~\cite{p}. 
With this remark, we finish the physical discussion of replicas in the context of GNJL models. 
In the next subsection, in order to evaluate the scale-dependent constituent quark mass $\tilde m$,
we resume this discussion within the effective-action approach.

\subsection{A correspondence between the GNJL and the effective-action approaches}\label{IIB}

From now on, we work in Euclidean space and 
denote $\tilde m$ as just $m$, while reserving the symbol $M$ for the heavy-quark current mass. 
In Ref~\cite{AntRib}, we established the correspondence between the GNJL and effective-action approaches. 
This correspondence becomes exact in the heavy-quark limit, which is the limit where \emph{both 
the Gaussian and the minimal-area (squared) approximations become exact}. For lower current quark masses, both approaches have 
to be supplemented by phenomenological-kernel ans\"atze. For the effective-action formalism this necessity is clear: 
When we depart from the heavy-quark limit, that is, from small Wilson loops of sizes of the order of the correlation length $a$, we depart from the minimal area they span. This leads us  
from the minimal plaquette that supports one homogeneous flux to an ensemble of correlated plaquettes, each having its 
own constant flux, which needs not to be the same across the neighborhood. Therefore, to establish a 
correspondence between the GNJL and the effective-action approaches, \emph{away} from the heavy-quark limit, where they are proven equivalent,
is the same as to establish a correspondence between their respective phenomenological extensions for lower quark masses. 
To this end, and as a first step, let us revisit the heavy-quark limit, where the sizes of quark trajectories, being of 
the order of the inverse current quark mass $1/M$, lie inside the circle of the size of the vacuum 
correlation length $a$. At distances smaller than $a$,
the non-Abelian field-strength tensor $F_{\mu\nu}^a$ can be treated as constant, yielding the gluon 
propagator
\begin{equation}
\label{Gluonprop}
D_{\mu\nu}^{ab}(x,y)\simeq\frac14 x_\alpha y_\beta\left<g^2F_{\alpha\mu}^a
\left(\frac{x+y}{2}
\right)F_{\beta\nu}^b\left(\frac{x+y}{2}\right)\right>.
\end{equation}
Furthermore, the correlation function of gluonic field strengths in this formula reads
\begin{equation}
\label{correlfun}
\left<g^2F_{\alpha\mu}^a\left(\frac{x+y}{2}
\right)F_{\beta\nu}^b\left(\frac{x+y}{2}\right)\right>=\bigl<G^2\bigr>\cdot\frac{\delta^{ab}}{N_c^2-1}
\cdot
\frac{1}{12}\bigl(\delta_{\alpha\beta}\delta_{\mu\nu}-\delta_{\alpha\nu}\delta_{\mu\beta}\bigr).
\end{equation}
Inserting Eqs.~(\ref{Gluonprop}) and (\ref{correlfun}) into Eq.~(\ref{njl2}) and introducing an 
antisymmetric tensor current 
$J_{\mu\nu}^a\equiv\frac12(j_\mu^a x_\nu-j_\nu^a x_\mu)$, we obtain the following GNJL action:
\begin{equation}
\label{NJL1}
S=\int_x \bar\psi(\gamma_\mu\partial_\mu+M)\psi+\frac{\bigl<G^2\bigr>}{96(N_c^2-1)}
\int_{x,y} J_{\mu\alpha}^a(x){\,}J_{\nu\beta}^b(y)\cdot \delta^{ab}
\bigl(\delta_{\alpha\beta}\delta_{\mu\nu}-\delta_{\alpha\nu}\delta_{\mu\beta}\bigr).
\end{equation}
This model has a confining kernel, which allows for replicas~\cite{fr}. Notice that, in Ref.~\cite{ne}, an 
extrapolation of the model to the chiral limit has been addressed.

Now, instead of solving the MGE corresponding to the GNJL model, Eq.~(\ref{NJL1}), 
one can map this model onto the one-loop 
quark effective action~\cite{AntRib}. To do so, one notices that the four-quark interaction in
 Eq.~(\ref{NJL1}) can be bosonized by introducing an auxiliary Abelian-like field 
${\cal A}_\mu^a(x)=\frac12 x_\nu n^a{\cal F}_{\nu\mu}$, where ${\cal F}_{\nu\mu}$ is a constant 
Abelian strength tensor, and $n^a$ is a constant unit color vector, $n^an^b=\delta^{ab}$. 
This observation
allows us to rewrite ${\rm e}^{-S}$ as
\be 
\label{newaction}
{\rm e}^{-S}=\left<{\rm e}^{-\int_x \bar\psi\left(\gamma_\mu D_\mu[{\cal A}]+M\right)\psi}\right>_{\cal F},
\ee
where 
\be
\label{Fmunu}
\bigl<\ldots\bigr>_{\cal F}\equiv
\frac{1}{(8\pi {\cal N})^3}
\left(\prod\limits_{\mu<\nu}^{}\int_{-\infty}^{+\infty}
d{\cal F}_{\mu\nu}\right){\rm e}^{-\frac{{\cal F}_{\mu\nu}^2}{16{\cal N}}}{\,} (\ldots),~~~~~ 
{\cal N}\equiv\frac{\bigl<G^2\bigr>}{48(N_c^2-1)},
\ee 
and the covariant derivative corresponding to the auxiliary field ${\cal A}_\mu^a$ reads
$D_\mu[{\cal A}]=\partial_\mu+i{\cal A}_\mu^aT^a$. The heavy-quark condensate is given by the tadpole 
diagram as 
\be
\label{quarkcond1}
\bigl<\bar\psi\psi\bigr>_{\rm heavy}=-\left<{\rm tr}{\,}\frac{1}{\gamma_\mu D_\mu[{\cal A}]+M}
\right>_{\cal F}.
\ee
It can be obtained upon the differentiation over $M$ of the one-loop effective action of a heavy quark 
in the auxiliary field ${\cal A}_\mu^a$:
\be
\label{quarkcond2}
\bigl<\bar\psi\psi\bigr>_{\rm heavy}=-\frac{\partial}{\partial M}{\,}\left<{\rm tr}{\,}\ln\left(
\gamma_\mu D_\mu[{\cal A}]+M\right)\right>_{\cal F}.
\ee
This expression is defined up to an overall normalization constant, which can be fixed upon 
a comparison with the 
free case, where the quark condensate vanishes. To divide the free case out means to perform the 
following substitution:
\be
\label{b7}
{\rm tr}{\,}\ln\left(
\gamma_\mu D_\mu[{\cal A}]+M\right) \rightarrow {\rm tr}{\,}\ln{\,}\frac{
\gamma_\mu D_\mu[{\cal A}]+M}{\gamma_\mu \partial_\mu+M}.\ee
We can now use the fact that the effective action of a quark in the 
field ${\cal A}_\mu^a$ is known in an analytic form (see, for example, the two reviews quoted in Ref.~\cite{WL}).
This observation yields 
\be\label{b8}
{\rm tr}{\,}\ln{\,}\frac{
\gamma_\mu D_\mu[{\cal A}]+M}{\gamma_\mu \partial_\mu+M}=-2N_{\rm f}\cdot {\rm tr}{\,}(T^aT^a)\cdot 
\int_0^\infty ds{\,}\frac{{\rm e}^{-M^2s}}{(4\pi s)^2}\left[abs^2\cot(as)\coth(bs)-1\right].
\ee
Here $s$ is the Schwinger proper time, 
${\rm tr}{\,}(T^aT^a)=C_{\rm f}\cdot{\rm tr}{\,}\hat 1_{N_c\times N_c}=(N_c^2-1)/2$, 
and $C_{\rm f}=(N_c^2-1)/(2N_c)$ is the quadratic Casimir operator of the fundamental representation 
of the SU($N_c$)-group. Furthermore, $a$ and $b$ can be expressed through the electric and magnetic 
fields corresponding to the Abelian strength tensor ${\cal F}_{\mu\nu}$ as 
\begin{eqnarray*}
a^2&=&\frac12\left[{\bf E}^2-{\bf H}^2+\sqrt{({\bf E}^2-{\bf H}^2)^2+4({\bf E}\cdot{\bf H})^2}\right],\\ b^2&=&\frac12\left[{\bf H}^2-{\bf E}^2+\sqrt{({\bf E}^2-{\bf H}^2)^2+4({\bf E}\cdot{\bf H})^2}\right].
\end{eqnarray*}
The effective action represents the sum of infinitely many diagrams with one quark loop and various numbers of 
external ${\cal A}_\mu^a$-lines. Due to the factor ${\rm e}^{-M^2s}$, one can readily see that the parameter 
of the expansion of the effective action 
in the number of external lines is $\bigl<G^2\bigr>/M^4$. Therefore, in the heavy-quark 
limit under study, the leading contribution is produced by the diagram with only two external lines.
One can further bring together Eqs.~(\ref{newaction})-(\ref{b8}) and proceed, when integrating over the 
${\cal F}_{\mu\nu}$-components in Eq.~(\ref{Fmunu}), to the 6-dimensional spherical coordinates. 
Using the value of the 6-dimensional solid angle, $\Omega_6=\pi^3$, one arrives at 
\begin{equation}
\label{he}
\left<\bar\psi\psi\right>_{\rm heavy}=-\frac{4N_c N_{\rm f}}{(8\pi C)^3}\cdot\frac{1}{(4\pi)^2}
\cdot\frac{\pi^3}{3}\cdot M\int_0^\infty ds{\,} {\rm e}^{-M^2s}\int_{0}^{1/s} dB B^7{\rm e}^{-
\frac{B^2}{8C}},
\end{equation}
where $C=\frac{\left<G^2\right>}{96N_c}$, and $B$ is the radial coordinate in the 6-dimensional space. 
In the heavy-quark limit at issue, the factor ${\rm e}^{-M^2s}$ filters out all proper times but those for
which $s^{-1}\gtrsim M^2$. For this reason, the upper limit in the $B$-integral can be replaced 
by $+\infty$, thereby decoupling the $s$- and the $B$-integrals from each other. 
This yields the heavy-quark condensate
\begin{equation}
\label{hea}
\left<\bar\psi\psi\right>_{\rm heavy}=-\frac{2N_c N_{\rm f}C}{\pi^2 M},
\end{equation}
which reproduces correctly the known result of QCD sum rules~\cite{svz}. By calculating the $B$-integral 
exactly as 
$$
\int_{0}^{1/s} dB B^7{\rm e}^{-\frac{B^2}{8C}}=3(8C)^4-\frac{4C}{s^6}
\left(\vphantom{\int}1+24Cs^2\left[1+16Cs^2\left(1+8Cs^2\right)\right]\right)\exp\left(-1/(8Cs^2)\right),
$$
one can see that corrections to Eq.~(\ref{hea}), arising from the entanglement of the two 
integrals in Eq.~(\ref{he}), are suppressed by an exponentially small factor 
${\cal O}\bigl({\rm e}^{-\frac{M^4}{8C}}\bigr)$. 

Let us now proceed towards the limit of small current quark masses. To this end, we model excitations of 
the minimal area by extending the gauge field as  
\be\label{subst}
A_\mu^a\to{\cal A}_\mu^a=A_\mu^a-\frac1g\sigma_\mu^a.
\ee  
Let us define the propagator 
$\bigl<g^2 \frac1g\sigma_\mu^a(x)\frac1g\sigma_\mu^a(y)\bigr>$  as $K_{\mu\nu}(x-y)$.
Then, this correction will be equivalent, in the lowest-loop approximation, to 
an extension of the action  
$\int_x\left[\bar\psi(\gamma_\mu D_\mu+m)\psi\right]$ by the current$\times$current interaction 
through the kernel $K_{\mu\nu}(x)$:
\be
\label{msaction} 
S=\int_x\left[\bar\psi(\gamma_\mu D_\mu+m)\psi\right]+
\int_{x,y} j_\mu^a(x)K_{\mu\nu}(x-y)j_\nu^a(y).
\ee
Here $D_\mu=\partial_\mu-igT^aA_\mu^a$ is the covariant derivative with $A_\mu^a$ yielding the minimal-area law. 
We furthermore assume $K_{\mu\nu}(x)$ to be long-range. 
Then, the current$\times$current interaction mediated 
by the long-range kernel $K_{\mu\nu}(x)$ should be thought as the generator of longitudinal string vibrations, 
which are not accounted for by the minimal-area law alone.
Our aim is to demonstrate how a long-range interaction kernel 
$K_{\mu\nu}(x)$ can be absorbed into the scale-dependence of the constituent quark mass $m$, 
as it was the case for the GNJL model -- see Eq.~(\ref{massdep}).
We note once again that, for heavy quarks whose trajectories fit into the circle of diameter $a$, 
the constituent quark mass goes over to the current mass $M$. Instead, light quarks acquire their 
constituent mass primarily in the course of their motion at distances $\gtrsim a$. This observation 
leads us to the natural assumption that constituent quark masses should not depend on a local point at the quark 
trajectory, but rather on the global characteristic of the trajectory -- in this case, its mean size $d$. This 
assumption is translated to the world-line formalism by means of the proper-time dependence 
of the constituent mass: $m=m(s)$. 

To see this, let us use the known world-line representation of the quartic self-interaction of a matter field.
In the simplest case, when the role of $j_\mu^a$ is played by the scalar density 
of a complex-valued Abelian Higgs field, this representation, reviewed in Appendix~A, involves an auxiliary scalar 
field $\sigma(x)$. This field is used to disentangle the 
quartic self-interaction via the Hubbard--Stratonovich transformation. In the actual case under study,
with the non-Abelian vector current $j_\mu^a$ reinstated,
this $\sigma$-field acquires a Lorentz and a color indices, that is becomes $\sigma_\mu^a$. 
A counterpart of Eq.~(\ref{A3}), $\int {\cal D}\psi{\cal D}\bar\psi {\rm e}^{-S}\simeq{\rm e}^{-\Gamma[A_\mu^a]}$,
defines an effective action $\Gamma[A_\mu^a]$. Its representation analogous to Eq.~(\ref{A4}) reads:
\be
\label{effectivea1}
\Gamma[A_\mu^a]=\int {\cal D}\sigma_\mu^a{\,} 
{\rm e}^{-\frac14\int_x \sigma_\mu^a(K^{-1})_{\mu\nu}\sigma_\nu^a}{\,}
\Gamma[A_\mu^a,\sigma_\mu^a],
\ee
where $\Gamma[A_\mu^a,\sigma_\mu^a]=-{\rm tr}{\,}\ln\left[\gamma_\mu(D_\mu+iT^a\sigma_\mu^a)+m\right]$.
Comparing $\Gamma[A_\mu^a,\sigma_\mu^a]$ with $\Gamma[A_\mu,\sigma]$, we see the main difference of the 
fermionic case at issue from the bosonic one of the Abelian Higgs model: 
Because in the fermionic case the kinetic term is 
first-order, 
the field $\sigma_\mu^a$ is included directly to $D_\mu$, and not to $D_\mu^2$ as in the bosonic case.
Notably, in the fermionic case, the field $\sigma_\mu^a$ \emph{just extends} the gauge field $A_\mu^a$ 
according to Eq.~(\ref{subst}). In terms of such an extended gauge field
and its strength tensor ${\cal F}_{\mu\nu}^a=\partial_\mu 
{\cal A}_\nu^a-\partial_\nu {\cal A}_\mu^a+gf^{abc}
{\cal A}_\mu^b{\cal A}_\nu^c$, 
a fermionic counterpart of Eq.~(\ref{A5}) reads
\be
\label{effectivea2}
\Gamma[A_\mu^a,\sigma_\mu^a]=-2N_{\rm f}V
\int_0^\infty\frac{ds}{s}{\rm e}^{-m^2s}\int_P {\cal D}z_\mu\int_A {\cal D}\psi_\mu 
{\rm e}^{-\int_0^s d\tau\left(\frac14\dot z_\mu^2+\frac12\psi_\mu\dot\psi_\mu\right)}{\,}
W[z_\mu,\psi_\mu,{\cal A}_\mu^a].
\ee
Here
\be
\label{Wilson1}
W[z_\mu,\psi_\mu,{\cal A}_\mu^a]={\,}{\rm tr}{\,}{\cal P}{\,}\exp\left[ig\int_0^s d\tau T^a
({\cal A}_\mu^a\dot z_\mu-\psi_\mu\psi_\nu {\cal F}_{\mu\nu}^a)\right],
\ee
$V$ is the 4D volume occupied by the system, $P$ and $A$ stand, respectively, for the periodic and the 
antiperiodic boundary conditions $\int_P^{}\equiv\int_{z_\mu(s)=z_\mu(0)}^{}$ and 
$\int_A^{}\equiv\int_{\psi_\mu(s)=-\psi_\mu(0)}^{}$, and the 
trajectories $z_\mu(\tau)$ obey the equation
$\int_0^s d\tau z_\mu(\tau)=0$, which constrains the center of a trajectory to the origin. Furthermore, 
the same lowest-loop approximation, used to derive  Eq.~(\ref{A6}), now yields 
\be
\int {\cal D}\sigma_\mu^a{\,} 
{\rm e}^{-\frac14\int_x \sigma_\mu^a(K^{-1})_{\mu\nu}\sigma_\nu^a}{\,}
W[z_\mu,\psi_\mu,{\cal A}_\mu^a]\simeq W[z_\mu,\psi_\mu,A_\mu^a]\cdot {\rm e}^{-C_{\rm f}
\oint dz_\mu\oint dz'_\nu K_{\mu\nu}(z-z')}.
\ee
Thus, similarly to the bosonic case, Eq.~(\ref{A6}), in the fermionic case we have also arrived 
at the self-interaction of the loop mediated by the kernel $K_{\mu\nu}(x)$.

\begin{figure}
\epsfig{file=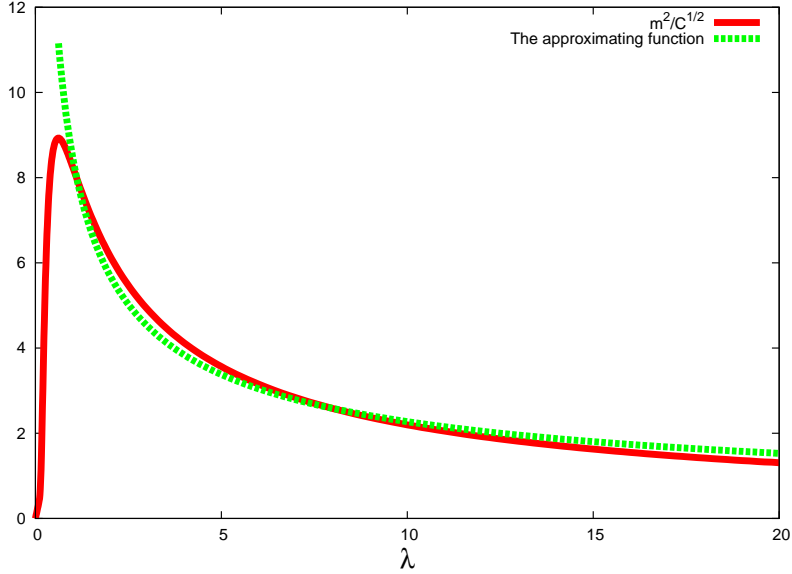, width=110mm}
\caption{\label{mm}The dimensionless constituent quark mass $m^2/\sqrt{C}$ as a function of the dimensionless proper 
time $\lambda=\sqrt{C}s$, Eq.~(\ref{massfunx}). The maximum of $m^2/\sqrt{C}$ is reached at 
$\lambda_{\rm extr}\simeq 0.62$. At $\lambda>\lambda_{\rm extr}$, is also shown 
the approximating function Eq.~(\ref{appr12}).}
\end{figure}

We are now ready to show that a 
long-range kernel can effectively be substituted by some proper-time 
dependent constituent quark mass $m(s)$. That is, we are seeking such a function $m(s)$ as 
to obey the equality 
\be
\label{equality1}
\int_P {\cal D}z_\mu
{\rm e}^{-\frac14\int_0^s d\tau\dot z_\mu^2-m^2(s)s}=\int_P {\cal D}z_\mu
{\rm e}^{-\frac14\int_0^s d\tau\dot z_\mu^2-C_{\rm f}
\oint dz_\mu\oint dz'_\nu K_{\mu\nu}(z-z')}.
\ee
By way of illustration, let us consider the simplest, quadratic, kernel 
$K_{\mu\nu}(x)=\frac{K_0^4}{2}\delta_{\mu\nu}x^2$, where $K_0$ is some parameter of the dimensionality of mass. 
Since for such a kernel one has 
$\oint dz_\mu\oint dz'_\nu K_{\mu\nu}(z-z')=-K_0^4\left(\oint dz_\nu z_\mu\right)^2$, the world-line 
integral on the right-hand side of Eq.~(\ref{equality1}) can again be calculated by introducing some auxiliary 
constant antisymmetric field $B_{\mu\nu}$ as follows:
$$
{\rm e}^{-C_{\rm f}\oint dz_\mu\oint dz'_\nu K_{\mu\nu}(z-z')} = \frac{1}{(8\pi C)^3}\left(\prod\limits_{\mu<\nu}^{}
\int_{-\infty}^{+\infty}
dB_{\mu\nu}\right) {\rm e}^{-\frac{B_{\mu\nu}^2}{16C}-\frac12B_{\mu\nu}\int_0^s d\tau z_\mu\dot z_\nu},
$$
where for the rest of this section
\be
\label{C0}
C\equiv C_{\rm f}K_0^4.
\ee 
The resulting world-line integral is given by the bosonic Euler--Heisenberg--Schwinger Lagrangian:
\be
\int_P {\cal D}z_\mu
{\rm e}^{-\int_0^s d\tau\left(\frac14\dot z_\mu^2+\frac12B_{\mu\nu}z_\mu\dot z_\nu\right)}=
\frac{1}{(4\pi s)^2}\cdot \frac{abs^2}{\sin(bs){\,}\sinh(as)}.
\ee
Expanding this Lagrangian up to the leading nontrivial term as
$$
\frac{abs^2}{\sin(bs){\,}\sinh(as)}\simeq 1-\frac{s^2}{6}\sum\limits_{\mu<\nu}^{}B_{\mu\nu}^2
$$ 
and proceeding in the $B_{\mu\nu}$-integration to 
the 6-dimensional spherical coordinates, we obtain
$$
\frac{1}{(4\pi s)^2}\cdot {\rm e}^{-m^2(s)s}~ \simeq\frac{1}{(4\pi s)^2}\cdot\frac12\int_0^{1/(8Cs^2)}
dy{\,}y^2\left(1-\frac{4Cs^2}{3}y\right){\rm e}^{-y},
$$
where $y\equiv(8C)^{-1}\sum\limits_{\mu<\nu}^{}B_{\mu\nu}^2$ is the radial integration variable 
in the 6-dimensional space.
Finally, introducing instead of $s$ a dimensionless proper time 
$\lambda\equiv\sqrt{C}s$, we obtain $m^2(\lambda)$ in the units of $\sqrt{C}$:
\be
\label{massfunx}
\frac{m^2(\lambda)}{\sqrt{C}}=-\frac{1}{\lambda}\cdot\ln\left[\frac12\int_0^{1/(8\lambda^2)}
dy{\,}y^2\left(1-\frac{4\lambda^2}{3}y\right){\rm e}^{-y}\right]\equiv f(\lambda).
\ee
This function is plotted in Fig.~\ref{mm}. It has a maximum at 
$\lambda_{\rm extr}\simeq0.62$, above which $m^2$ steadily falls off and can be approximated by the function 
\begin{equation}
\label{appr12}
f_{\rm appr}(\lambda)=\frac{8.48\sqrt{C}}{\lambda^{0.57}}.
\end{equation} 
Thus, at sufficiently large proper times relevant for the spontaneous 
breaking of chiral symmetry, the effective constituent quark mass squared approximately falls off as 
$$m^2(s)\sim\frac{1}{s^{0.57}}.$$

We can now obtain the effective constituent mass as a function of the physical mean size $d$ of a trajectory.
It has been shown in Ref.~\cite{AntRib}
that the classical trajectories of a confined light quark are circles, so that the mean size $d$ 
is the diameter of a circle.
We notice that, for an $s$-dependent mass $m(s)$ at issue, the length of the quark trajectory $L$
is represented by the integral $L(s)=\int_0^s m(\tau)d\tau$. Using the function $f(\lambda)$ introduced in 
Eq.~(\ref{massfunx}), we can write the length $L$ 
as an integral over the dimensionless proper time $\lambda$ as 
\begin{equation}
\label{Lx}
L(\lambda)=C^{-1/4}\int_0^\lambda d\lambda'\sqrt{f(\lambda')}.
\end{equation}

\begin{figure}
\epsfig{file=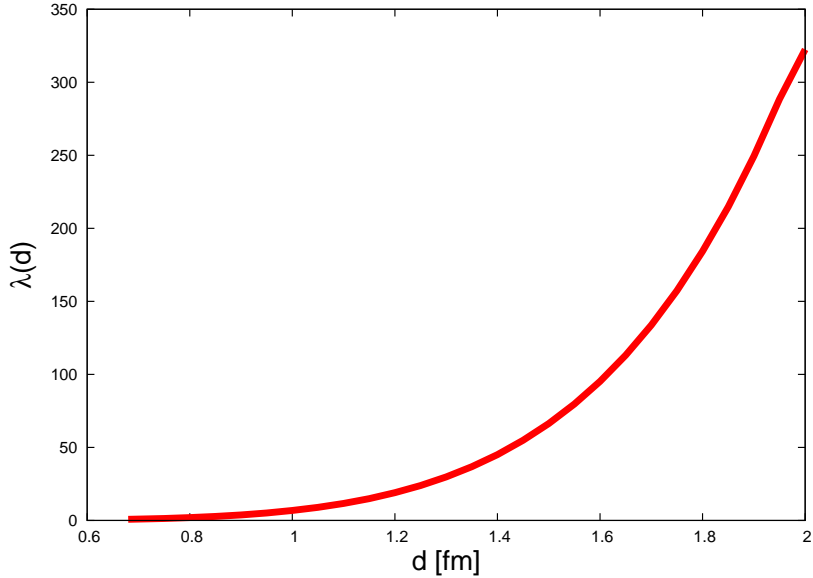, width=110mm}
\caption{\label{xd}The dimensionless proper 
time $\lambda=\sqrt{C}s$ as a function of the mean size $d$ of a trajectory.}
\end{figure}

\begin{figure}
\epsfig{file=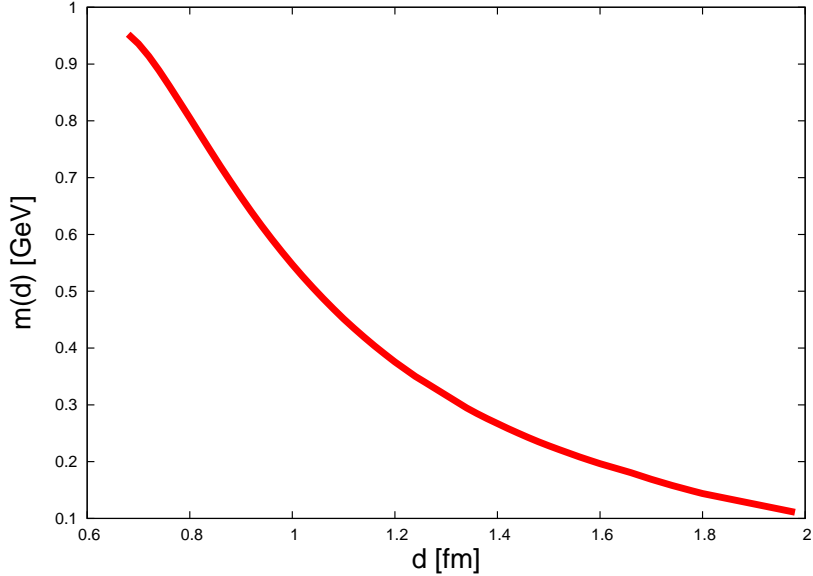, width=110mm}
\caption{\label{Md}The effective constituent quark mass $m$
as a function of the mean size $d$ of a trajectory.}
\end{figure}

\noindent
We can furthermore use the result of Ref.~\cite{AntRib} that a light-quark trajectory has Hausdorff dimension 4, that is its length grows as the fourth power of its mean size, $L\propto d^4$. Since the minimal 
mean size of a trajectory, for which the quark is still confined, is $d_{\rm min}=2a$, 
we obtain the following formula for the length of a trajectory in terms of 
its mean size:
\begin{equation}
\label{Ld}
L(d)=L(d_{\rm min})\cdot\left(\frac{d}{d_{\rm min}}\right)^4,~~~ {\rm where}~~~ L(d_{\rm min})=
\pi\cdot d_{\rm min}.
\end{equation}
Equating to each other the two expressions for the length, Eqs.~(\ref{Lx}) and (\ref{Ld}), we can obtain the dimensionless proper time in terms of the mean size of the loop, $\lambda(d)$. We choose the range of distances 
$d\in[d_{\rm min},2{\,}{\rm fm}]$, use 
the commonly accepted 
value of the vacuum correlation length~\cite{wq, en, smalla} $a=0.34{\,}{\rm fm}$, and evaluate the 
parameter $C$, Eq.~(\ref{C0}), with the strength 
of the kernel $K_0=300{\,}{\rm MeV}$ (cf. the second reference in~\cite{fr}). We notice 
further that $\lambda(d)$ 
is a monotonically increasing function, and that already $\lambda(d_{\rm min})\simeq0.79$ exceeds 
the position of the maximum in Fig.~\ref{mm}, $\lambda_{\rm extr}\simeq0.62$. This fact 
facilitates the procedure of the numerical solution of the equation $L(\lambda)=L(d)$.
Namely, instead of using in the length-functional, Eq.~(\ref{Lx}), 
an integral representation of the function $f$, Eq.~(\ref{massfunx}), we use the approximating 
function $f_{\rm appr}$, Eq.~(\ref{appr12}). In this way, we obtain the function $\lambda(d)$ and 
plot it in Fig.~\ref{xd}. Substituting now $\lambda$ in Eq.~(\ref{massfunx}) by this function $\lambda(d)$, we obtain also 
the effective constituent quark mass in terms of the mean size of a trajectory. Such a function
$m(d)$ is plotted in Fig.~\ref{Md}.
In particular, we find that $m(d)$ acquires its commonly accepted phenomenological value of 300~MeV at 
$d\simeq 1.30{\,}{\rm fm}$. The maximum value, which the constituent mass acquires, is the one 
at $d=d_{\rm min}$, that is $m_{\rm max}=953{\,}{\rm MeV}$.  
\emph{Thus, we have shown that, for a given 
confining kernel $K_{\mu\nu}(x)$, we also have, like in the GNJL case, a running mass  $m(d)$, this time a function of the loop size $d$}. 
In particular, the constituent quark mass $m_{\rm max}$ defines the scale for the highest possible replica, arising when
the loop is shrunk down to the size of $d_{\rm min}$. We notice finally that, in addition to $d_{\rm min}$, there 
always exists some limiting maximum value $d_{\rm max}$. It is associated with the string-breaking, as 
at $d>d_{\rm max}$ the system becomes unstable due to the quark-antiquark pair creation.
Lattice data and analytic models suggest some value around 2~fm~\cite{sb} for the string-breaking distance. 

We therefore take the combined theoretical evidence for the existence of replicas, coming from both 
the GNJL and the effective-action formalisms, as a hint that they might be realized in Nature. However, 
in order for this to have a chance to happen, we still have to evaluate, through the use of the Tolman--Oppenheimer--Volkoff 
equation, the stability condition for a replica-filled domain versus the gravity field it might produce. 
To do this, we need to know the equation of state for such macroscopic objects. This will be the subject of the 
next section. Finally, in section~IV, we extend the effective-action approach to evaluate the leading gravitational correction 
to the quark condensate. There, it will be shown that for distances larger than the hadronic scale, 
gravitational corrections 
to the flat-space results are negligible. This fact allows us to use the standard flat-space value of the chiral condensate 
to study the equation of state of such macroscopic domains.

\section{The energy density and macroscopic properties of a replica-filled domain}

Let us proceed with the evaluation of the full internal-energy density $\varepsilon(T)$ of a macroscopic 
domain whose quantum states are built on top of a replica state. The full internal-energy density of such an object is given
by the sum of the replica-vacuum contribution and the contribution produced by hadronic 
excitations above that vacuum (clearly, staying in the lowest-energy domain, we restrict ourselves to pionic excitations):
\begin{equation}
\label{-1}
\varepsilon(T)=\varepsilon_{\rm vac}(T)+\varepsilon_h(T).
\end{equation}
We consider first the vacuum contribution $\varepsilon_{\rm vac}(T)$. It is defined as a difference of the
internal-energy density of the excited vacuum, corresponding to a replica state inside the domain, 
and the internal-energy density of the unexcited vacuum outside the domain:
\begin{equation}
\label{0}
\varepsilon_{\rm vac}(T)=\varepsilon_R(T)-\varepsilon_0(T).
\end{equation}
The internal-energy density of the unexcited vacuum has the form
\begin{equation}
\label{1}
\varepsilon_0(T)=\frac14\left[-\frac{b}{32\pi^2}\left<G^2\right>_T+(m_u+m_d)\left<\bar\psi\psi\right>_T\right],
\end{equation}
where the one-loop $\beta$-function coefficient for $N_c=3$, $N_{\rm f}=2$ is $b=29/3$, and the 
approximation~\cite{1} $\left<\bar\psi\psi\right>\simeq\left<\bar uu\right>
\simeq\left<\bar dd\right>$ has been adopted. One can further use the known temperature-dependent 
gluonic and chiral condensates, which at temperatures $T\ll m_\pi$ of interest read~\cite{1}
\begin{equation}
\label{2}
\left<G^2\right>_T=\left<G^2\right>-\frac{24m_\pi^3T}{b} S_1\left(\frac{m_\pi}{T}\right),\quad
\left<\bar\psi\psi\right>_T=\left<\bar\psi\psi\right>\left[1-\frac{3m_\pi T}{4\pi^2f_\pi^2}
S_1\left(\frac{m_\pi}{T}\right)\right].
\end{equation}
Here, $\left<G^2\right>$ and $\left<\bar\psi\psi\right>$ are the zero-temperature values of the condensates in the
unexcited vacuum, $f_\pi$ is the pion decay constant in the same vacuum, and 
we denote the sums entering virial expansions of the grand canonical ensembles of relativistic bosons as 
$$S_\nu(x)\equiv\sum\limits_{n=1}^{\infty}\frac{K_\nu(nx)}{n^\nu},$$
where $K_\nu$ is the MacDonald function. Inserting Eqs.~(\ref{2}) into Eq.~(\ref{1}), and using the 
Gell-Mann--Oakes--Renner relation~\cite{GMOR} $(f_\pi m_\pi)^2=-(m_u+m_d)\left<\bar\psi\psi\right>$, we 
obtain
\begin{equation}
\label{3}
\varepsilon_0(T)=\varepsilon_0(0)+\frac{3m_\pi^3T}{8\pi^2} S_1\left(\frac{m_\pi}{T}\right),
\end{equation}
where $\varepsilon_0(0)$
is the internal-energy density of the unexcited vacuum at zero temperature. For the 
internal-energy density of the excited vacuum, in which the 
Gell-Mann--Oakes--Renner relation holds equally well (cf. the last reference in~\cite{MoreRep}), an expression similar to Eq.~(\ref{3}) reads:
\begin{equation}
\label{4}
\varepsilon_R(T)=\varepsilon_R(0)+\frac{3m_{\pi_R}^3T}{8\pi^2} S_1\left(\frac{m_{\pi_R}}{T}\right),
\end{equation}
where $m_{\pi_R}$ is the mass of the pion in the replica.
Equations~(\ref{0}), (\ref{3}), and (\ref{4}) yield 
\begin{equation}
\label{vac}
\varepsilon_{\rm vac}(T)=\varepsilon+\frac{3T}{8\pi^2}\left[m_{\pi_R}^3 S_1
\left(\frac{m_{\pi_R}}{T}\right)-
m_\pi^3 S_1\left(\frac{m_\pi}{T}\right)\right],
\end{equation}
where $\varepsilon\equiv\varepsilon_R(0)-\varepsilon_0(0)$.

\begin{figure}
\epsfig{file=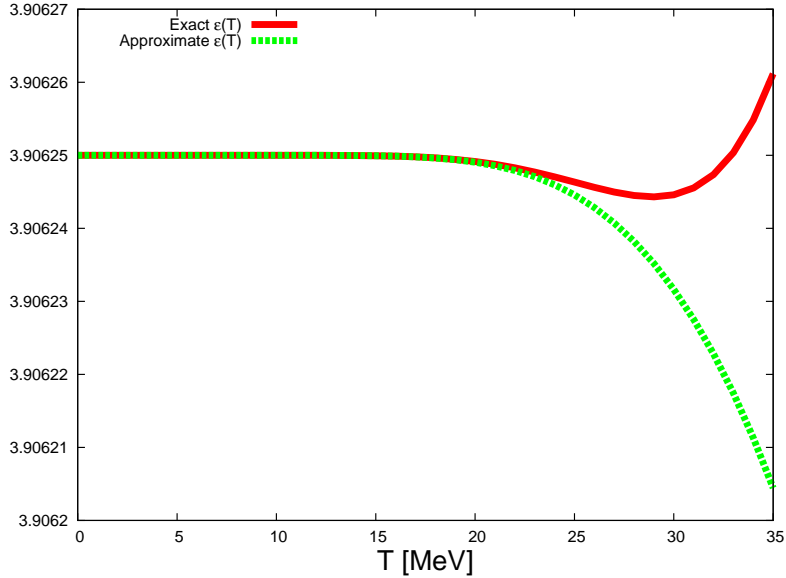, width=110mm}
\caption{\label{t}The full internal-energy density, $\varepsilon(T)\cdot 10^{-9}{\,}({\rm MeV})^4$, given by 
Eq.~(\ref{-1}), and its approximation, Eq.~(\ref{var}). 
The constant value, $\varepsilon(T)\cdot 10^{-9}=3.90625{\,}({\rm MeV})^4$, corresponds to 
$\varepsilon(T)\simeq \varepsilon=(250{\,}{\rm MeV})^4$, Eq.~(\ref{ep}).}
\end{figure}

We find now the hadronic contribution $\varepsilon_h(T)$ to the internal-energy density~(\ref{-1}).
By neglecting the outer temperature with respect to the temperature inside the domain, we neglect 
also the outer pressure. The pressure of the relativistic pionic gas inside the domain reads~\cite{1}:
\begin{equation}
\label{5}
p_h(T)=\frac{3m_{\pi_R}^2T^2}{2\pi^2} S_2\left(\frac{m_{\pi_R}}{T}\right). 
\end{equation}
The hadronic contribution to the internal-energy density can be found by the standard thermodynamics 
formula
\begin{equation}
\label{hadr}
\varepsilon_h(T)=T\frac{d p_h(T)}{dT}-p_h(T).
\end{equation}
Finally, according to Eq.~(\ref{-1}), 
the full internal-energy density is given by the sum of Eqs.~(\ref{vac}) and~(\ref{hadr}). 
Noticing that $m_{\pi_R}>m_\pi$~\cite{MoreRep}, we see that the leading correction  
to the formula $\varepsilon(T)\simeq\varepsilon$ at temperatures $T\ll m_\pi$ reads
\begin{equation}
\label{var}
\varepsilon(T)\simeq\varepsilon-\frac{3Tm_\pi^3}{8\pi^2}K_1\left(\frac{m_\pi}{T}\right).
\end{equation}
Higher corrections are getting more important with the increase 
of temperature. Using the values~\cite{MoreRep}
\be
m_{\pi_R}=250~\mbox{MeV}, \quad\langle\bar{\psi}\psi\rangle_R=-(100~\mbox{MeV})^3,
\quad\varepsilon\simeq(250{\,}{\rm MeV})^4, 
\label{values}
\ee
and the standard $\pi$-meson mass $m_\pi=140{\,}{\rm MeV}$, we evaluate numerically 
the internal-energy density~(\ref{-1}) 
along with its approximate expression~(\ref{var}), and plot these quantities in Fig.~\ref{t}. 
We observe, in particular, 
the complete numerical irrelevance of the hadronic contribution up
to the temperatures $\sim 25{\,}{\rm MeV}$. For this reason, in what follows we disregard the hadronic 
contribution
altogether, and approximate the internal-energy density of the domain by the constant value 
\begin{equation}
\label{ep}
\varepsilon(T)\simeq \varepsilon=(250{\,}{\rm MeV})^4.
\label{e250}
\end{equation}

It follows from Eq.~(\ref{norm}) that, with an exponential accuracy, a replica-filled domain occupying a sufficiently large volume $V^{(3)}$ is stable against the decays into the ground-state vacuum. The 
phenomenon providing this stability is the very strong correlations between pions in the replica. We notice also that, with the energy density (\ref{e250}), the mass of a replica-filled domain grows with the volume as $\varepsilon\times V^{(3)}$, so that the problem of stability with respect to the gravitational collapse becomes relevant. For a constant energy-density, a solution to the so-called Tolman--Oppenheimer--Volkoff equation,
which provides the metric of a spherically-symmetric object is well known~\cite{w}.
In Appendix~B, we briefly discuss this solution, which defines the maximal possible radius of the domain [see Eq.~(\ref{B8})] as
\begin{equation}
\label{radius}
R_G=\frac{1}{\sqrt{3\pi\varepsilon G}}\simeq 14{\,}{\rm km},
\end{equation}
where $G=6.7\cdot 10^{-39}{\,}{\rm GeV}^{-2}$ is the gravitational constant.
Notice that the obtained estimate for the maximum radius is of order of the typical radius of a 
neutron star, that is about 10~km. This analogy extends further to the rotating neutron stars (pulsars).
Namely, consider a domain of mass $M$ rotating with the angular velocity $\Omega$. Then, for an element 
of  matter of
mass $\mu$ situated at the equator of the domain (where the centrifugal force is maximal), the Newton's 
second law has
the form
$$\xi^2 G\frac{\mu M}{R_G^2}=\mu\Omega^2 R_G,$$
where the correcting factor $\xi\simeq 0.65$ accounts for the effects of General Relativity~\cite{gl}.
Noticing that $M=4\pi\varepsilon R_G^3/3$, we obtain from this Newton's equation the period of the domain's 
rotation
\begin{equation}
\label{period}
T_G=\frac{2\pi}{\Omega}\simeq 0.6{\,}{\rm msec}.
\label{Tpuls}
\end{equation}
Notice that periods of rotation of certain pulsars can be as small as milliseconds, that qualitatively agrees with the value~(\ref{Tpuls}).
 
We end this section with a brief discussion on the stability of the replica-filled domains
against the decays of their pionic excitations into photons.
Such anomalous decays appear suppressed compared to the similar conventional 
decays in the unexcited vacuum. To see this suppression, we 
use the Gell-Mann--Oakes--Renner relation~\cite{GMOR}, which yields the pion decay constant $f_\pi$. 
Since in the replica the chiral symmetry is ``less broken'' than in the unexcited vacuum, 
the absolute value of the chiral condensate in the replica is smaller, whereas the pion mass is larger.
Using the values of $m_{\pi_R}$ and $\langle\bar{\psi}\psi\rangle_R$ quoted in Eq.~(\ref{values}),
one can readily evaluate a two-photon decay width of a ``replicated'' pion $\pi_R$ as
\be
\Gamma(\pi_R\to\gamma\gamma)=\Gamma(\pi\to\gamma\gamma)\cdot
\frac{f_{\pi_R}^2}{f_\pi^2}=
\Gamma(\pi\to\gamma\gamma)\cdot\frac{\langle\bar{\psi}\psi\rangle_R}
{\langle\bar{\psi}\psi\rangle_0}{\,}\frac{m_\pi^2}{m_{\pi_R}^2}\simeq
0.17{\,}{\rm eV}, 
\label{wd} 
\ee 
whereas $\Gamma(\pi\to\gamma\gamma)\simeq 8{\,}{\rm eV}$.
The smallness of the calculated decay width 
$\Gamma(\pi_R\to\gamma\gamma)$ suggests that a replica-filled domain would appear to an external observer as a dark massive body
made of strongly correlated pions.

\section{A gravitational correction to the chiral condensate}

The estimate of the radius of a replica-filled domain, made in the previous section, employed the flat-space result for the vacuum energy-density. It is clear, however, that chiral symmetry breaking experiences influence of gravity and that, in general, the MGE and the Tolman--Oppenheimer--Volkoff equation are mutually 
coupled. On the other hand, the estimated radius $R_G$ of a stable domain constitutes a few kilometers, while the scale of chiral symmetry $r_\chi$ 
is as small as just a few fermi, that is 
many orders of magnitude smaller than $R_G$.
In other words, we have a well pronounced hierarchy of scales in the problem, and thus we anticipate that gravitational corrections to chiral symmetry are negligibly small.

In this section, we justify the above qualitative argument by evaluating
the leading correction acquired by the chiral condensate in the gravitational field of a replica-filled domain. 
Such a correction is expected to be of the form 
${\cal O}(\partial^2)/m^2$, where $m$ is the constituent quark mass, and 
the ${\cal O}(\partial^2)$-terms contain in particular the Ricci scalar 
${\cal R}$. Within the effective-action approach, 
the quark condensate stems from the formula $\left<\bar\psi\psi\right>=-
\frac{\partial}{\partial m}\left<\Gamma[A_\mu^a]\right>$. In the flat space, 
yet another derivation of Eq.~(\ref{he}) can be performed (cf. Appendix~A of Ref.~\cite{AntRib}), 
which is based on the one-loop effective action
$$\left<\Gamma[A_\mu^a]\right>=-\frac{2N_c N_{\rm f}}{(8\pi C)^3}\int_0^\infty\frac{ds}{s}{\rm e}^{-m^2s} 
\left(\prod\limits_{\mu<\nu}^{}\int_{-\infty}^{+\infty}
dB_{\mu\nu}\right){\rm e}^{-\frac{B_{\mu\nu}^2}{16C}}\times$$
$$\times\left\{
\int_{P}^{}{\cal D}z_\mu \int_{A}^{}{\cal D}\psi_\mu\exp\left[-\int_0^s d\tau\left(
\frac14\dot z_\mu^2+\frac12\psi_\mu\dot\psi_\mu+\frac{i}{2}B_{\mu\nu}z_\mu\dot z_\nu-
iB_{\mu\nu}\psi_\mu\psi_\nu\right)\right]-\frac{1}{(4\pi s)^2}\right\},$$
where again $C=\frac{\bigl<G^2\bigr>}{96N_c}$, and $B_{\mu\nu}$ is an auxiliary constant Abelian field.
As is shown in Ref.~\cite{WL}, the result of the functional integration in this formula is given by the
Euler--Heisenberg--Schwinger Lagrangian, which can further be Taylor-expanded in $s$:
\begin{equation}
\label{heat}
\left\{\cdots\right\}=\frac{1}{(4\pi s)^2}\cdot\sum\limits_{n=0}^{\infty} a_n s^n.
\end{equation}
That is, of course, just the heat-kernel expansion of the (logarithm of the) determinant of the Dirac 
operator of an
electron 
in the electromagnetic field $B_{\mu\nu}$. Each term of this expansion is described by the one-loop 
diagram with the corresponding number of external lines of the $B_{\mu\nu}$-field. In particular, the 
leading non-vanishing term, with the coefficient 
$a_2=\frac16 B_{\mu\nu}^2$, which corresponds to the diagram with two such lines, recovers  
Eq.~(\ref{he}), where $B=\sum\limits_{\mu<\nu}^{} B_{\mu\nu}^2$.

A distinguishing property of the heat-kernel expansion in the curved (Euclidean) 
space is that some of the diagrams with 
two external lines of the $B_{\mu\nu}$-field contribute not only to the 2-nd, but also to the 3-rd term in
Eq.~(\ref{heat}). That is because such diagrams additionally describe an interaction of the electron with 
the
gravitational field.
This interaction involves operators of dimensionality (mass)$^2$, such as the Ricci scalar ${\cal R}$, 
which are accompanied by an extra power of $s$. 
Explicitly, the coefficient $a_3$ in the presence 
of the gravitational field has the form~\cite{gr}
\begin{equation}
\label{a3}
a_3=-\frac{1}{360}\left[5{\cal R}B_{\mu\nu}B^{\mu\nu}-26{\cal R}^{\mu\nu}B_{\mu\lambda}
B_\nu{\,}^\lambda+2{\cal R}_{\mu\nu\lambda\rho}B^{\mu\nu}B^{\lambda\rho}+24(\nabla_\nu B^{\mu\nu})
(\nabla^\lambda B_{\mu\lambda})\right].
\end{equation}
Notice also that the measure of integration over the $B_{\mu\nu}$-field is changed in the curved space, 
where
$$\left(\prod\limits_{\mu<\nu}^{}\int_{-\infty}^{+\infty}
dB_{\mu\nu}\right){\rm e}^{-\frac{B_\lambda{\,}^\rho B^\lambda{\,}_\rho}{16C}}=
\frac{(8\pi C)^3}{(\det g^{\mu\nu})^{3/2}}.$$

Performing the $s$-integrations, we arrive then at the following intermediate expression for the quark 
condensate:
\begin{equation}
\label{ch1}
\left<\bar\psi\psi\right>=-\frac{N_cN_{\rm f}}{4\pi^2}\cdot\frac{(\det g^{\mu\nu})^{3/2}}{(8\pi C)^3}
\left(\prod\limits_{\mu<\nu}^{}\int_{-\infty}^{+\infty}
dB_{\mu\nu}\right){\rm e}^{-\frac{B_\lambda{\,}^\rho B^\lambda{\,}_\rho}{16C}}\left(\frac{a_2}{m}+
\frac{a_3}{m^3}\right),
\end{equation}
where $a_2=\frac16B_{\mu\nu}B^{\mu\nu}$, while $a_3$ is given by Eq.~(\ref{a3}). In particular, for 
$a_3\equiv 0$ and $g^{\mu\nu}=\delta^{\mu\nu}$, Eq.~(\ref{ch1}) reproduces Eq.~(\ref{hea}).
The $B_{\mu\nu}$-integration in Eq.~(\ref{ch1}) can be performed through the formula
$$\frac{(\det g^{\mu\nu})^{3/2}}{(8\pi C)^3}
\left(\prod\limits_{\mu<\nu}^{}\int_{-\infty}^{+\infty}
dB_{\mu\nu}\right){\rm e}^{-\frac{B_\lambda{\,}^\rho B^\lambda{\,}_\rho}{16C}}B^{\alpha\beta}B^{\gamma\sigma}=
4C(g^{\alpha\gamma}g^{\beta\sigma}-g^{\alpha\sigma}g^{\beta\gamma}).$$
We furthermore proceed to the Minkowski space, since the metric of the domain to be eventually used is 
Minkowskian.
There, due to the constancy and antisymmetry of $B_{\mu\nu}$,
one has $\nabla_\nu B^{\mu\nu}=B^{\mu\nu}\partial_\nu \ln\sqrt{-g}$, where $g=\det g_{\mu\nu}$.
Finally, contracting the indices, we obtain for the quark condensate the following result:
\begin{equation}
\label{le}
\left<\bar\psi\psi\right>=\left<\bar\psi\psi\right>_0\left[1+\frac{1}{60m^2}\left({\cal R}-
6\cdot g^{\mu\nu}\partial_\mu \ln\sqrt{-g}\cdot \partial_\nu \ln\sqrt{-g}{\,}\right)\right],
\end{equation}
where $\left<\bar\psi\psi\right>_0$ is the condensate in the flat space. 

Equation~(\ref{le}) can now be used to evaluate the quark condensate 
in the gravitational field of a replica-filled spherical domain of 
a constant energy-density. We start
with the inner part of the domain and assume that it has the maximum possible radius~(\ref{radius}).
Now, inserting Eq.~(B7) into Eq.~(B9), we obtain the 
following expression for the temporal component of the metric inside the domain:
\be
\label{g00}
g_{00}(r)=\frac19\cdot\exp\left[-\frac12\int\limits_{z}^{8/9}\frac{dz'}{1-z'}
\left(1+\frac{3\sqrt{1-z'}-1}{1-\sqrt{1-z'}}\right)\right],
\ee
where $z\equiv 8\pi\varepsilon Gr^2/3$ is such that $z<8/9$. Using also the other components of the metric, namely 
\be
g_{11}=\frac{1}{z-1},~~~ g_{22}=-r^2,~~~ g_{33}=-r^2\sin^2\theta ,
\ee
we have for the corresponding term in Eq.~(\ref{le}):
\be
g^{\mu\nu}\partial_\mu \ln\sqrt{-g}\cdot \partial_\nu \ln\sqrt{-g}=
\left(z-1\right)
\left[\frac{d}{dr}{\,}\ln\left(r^2\sqrt{g_{00}g_{11}}\right)\right]^2-\frac{\cot^2\theta}{r^2}.
\ee
The differentiation in this formula is straightforward and yields 
\begin{equation}
\label{t2}
g^{\mu\nu}\partial_\mu \ln\sqrt{-g}\cdot \partial_\nu \ln\sqrt{-g}=\frac{1}{r^2}\left[
\frac{3-z+\sqrt{1-z}}{z-1}-
\cot^2\theta\right].
\end{equation}
Due to the spherical symmetry of the domain, we are interested only in the radial dependence of the 
gravitational 
correction to the quark condensate, while averaging out its angular dependence. That can be done by 
noticing 
that the azimuthal angle enters the invariant 
volume element as $d^4x\sqrt{-g}\propto d\theta\sin^2\theta$, thereby defining  
the following normalized angular average:
$\left<\cdots\right>_\theta=\frac{2}{\pi}\int_0^\pi d\theta\sin^2\theta(\cdots)$.
In particular, the so-averaged angular-dependent term in Eq.~(\ref{t2}) yields just unity:
$\left<\cot^2\theta\right>_\theta=1$. 

The Ricci scalar ${\cal R}$ entering Eq.~(\ref{le}) can be expressed through the functions 
$a(r)=\ln g_{00}$ and $b(r)=\ln(-g_{11})$ as follows~\cite{w, ll}:
\be
\label{Ricci}
{\cal R}={\rm e}^{-b}\left[\frac{d^2a}{dr^2}+\frac2r\cdot\frac{d(a-b)}{dr}+\frac12\left(\frac{da}{dr}
\right)^2-\frac12\frac{da}{dr}\cdot \frac{db}{dr}+\frac{2}{r^2}\right]-\frac{2}{r^2}.
\ee
A straightforward calculation yields for it a simple result:
\be
\label{Ricci2}
{\cal R}=\frac{12(1-z)}{r^2}.
\ee
The full $\theta$-averaged correction to the quark condensate, Eq.~(\ref{le}), thus reads
\begin{equation}
\label{in}
\frac{1}{60m^2}\left(
{\cal R}-6\cdot\left<g^{\mu\nu}\partial_\mu \ln\sqrt{-g}\cdot 
\partial_\nu \ln\sqrt{-g}{\,}\right>_\theta\right)
=\frac{1}{10(mr)^2}
\left(4+\frac{1}{\sqrt{1-z}}+\frac{2}{1-z}-2z\right).
\end{equation}
The obtained correction to
the absolute value of the quark condensate, $|\left<\bar\psi\psi\right>_0|$,
is positive-definite, and it gets smaller than 1 when the following condition is met:
\begin{equation}\label{ineq1}
\label{bbn87}
\frac{1}{z}\left(4+\frac{1}{\sqrt{1-z}}+\frac{2}{1-z}\right)-2<\frac{15m^2}{4\pi\varepsilon G}.
\end{equation}
To numerically evaluate the obtained correction in the chiral limit, we use 
the internal-energy density~(\ref{ep}), as well as 
the values of the constituent quark mass $m=300{\,}{\rm MeV}$ and 
the gravitational constant $G=6.7\cdot 10^{-39}{\,}{\rm GeV}^{-2}$.
We obtain for the right-hand side of inequality~(\ref{bbn87}):
$15m^2/(4\pi\varepsilon G)\simeq4.1\cdot 10^{39}$. Seeking distances $r$ at which the 
obtained correction to the condensate can become relevant, we get from the definition of $z$ and Eq.~(\ref{ineq1})
\begin{equation}
\label{in1}
r<\frac{\sqrt{7/10}}{m}\simeq0.55{\,}{\rm fm}.
\end{equation}
This entails the following correction to the quark condensate:
\be
\langle\bar\psi\psi\rangle=\langle\bar\psi\psi\rangle_0 \left[1+{\cal O}((r_{\chi}/R_G)^2)\right],
\ee
where $R_G$ is given by Eq.~(\ref{radius}) and $r_\chi=1/m$, with $m$ being the effective 
constituent quark mass, which appears due to chiral symmetry breaking. One can thus see that, as was anticipated, inside the domain, the effect of its gravitational field on the chiral condensate is negligibly small.

Outside the domain $(r>R_G)$, the metric is Schwarzschild (cf. Appendix~B), therefore ${\cal R}=0$. 
The remaining  term in Eq.~(\ref{le}) reads for this metric:
\be
\label{jnl987}
g^{\mu\nu}\partial_\mu \ln\sqrt{-g}\cdot \partial_\nu \ln\sqrt{-g}=
\frac{1}{r^2}\left[4\left(\frac{r_g}{r}-1
\right)-\cot^2\theta\right],
\ee
where $r_g\equiv 2 G M$ is the Schwarzschild radius, with $M$ being the full energy of the domain -- see Appendix~B.
Once averaged over $\theta$, Eq.~(\ref{jnl987}) yields the following positive-definite correction to $|\left<\bar\psi\psi\right>_0|$: 
\begin{equation}
\label{out}
\frac{1}{60m^2}\left(
{\cal R}-6\cdot\left<g^{\mu\nu}\partial_\mu \ln\sqrt{-g}\cdot \partial_\nu \ln\sqrt{-g}{\,}\right>_\theta
\right) 
=\frac{1}{10(mr)^2}\left[4\left(1-\frac{r_g}{r}\right)+1\right].
\end{equation}
Noticing that $r_g/R_G=8/9$, we see that $\left[4\left(1-\frac{r_g}{r}\right)+1\right]$ grows 
from the value $13/9$ at $r=R_G$ to the asymptotic value 5 at $r\to\infty$. Therefore, we can estimate the 
obtained correction as 
\begin{equation}
\label{out1}
\frac{1}{10(mr)^2}\left[4\left(1-\frac{r_g}{r}\right)+1\right]\le\frac{5}{10(mR_G)^2}=
\frac{3\pi\varepsilon G}{2m^2}\simeq1.4\cdot 10^{-39}.
\end{equation}
Such a numerical smallness of the result originates, of course, from the value 
of the gravitational constant $G$. We conclude that, outside 
the domain, the effect of its gravitational field on the chiral condensate is negligible as well.

\section{Summary}

In this paper, we considered the possibility of formation in the early Universe of macroscopic domains 
filled with coherently excited $\pi$-mesons. 
At temperatures below 25~MeV, the energy density inside the domains is primarily defined by the energy gap 
over the physical QCD vacuum of the first excited coherent state (the so-called first replica) in the GNJL model. 
Instead, the pionic excitations above the replica are shown to be 
suppressed at such temperatures. 
In this way, using the constant energy-density~(\ref{ep}), we obtained 
the maximum possible radius of the
domain, Eq.~(\ref{radius}), which turns out to be similar to the radius of a neutron star. Furthermore, for
rotating domains, the period of their rotation~(\ref{period}) is also similar to that of some class of pulsars.

Since our approach is based on the energy density in the flat space, we have checked its 
self-consistency by demonstrating a smallness of the leading correction to the flat-space chiral 
condensate appearing in the gravitational field of the domain. To this end, 
we have first illustrated the correspondence between the GNJL and the effective-action descriptions of
spontaneous chiral symmetry breaking. Then, using the effective-action approach in the curved space, 
we derived the leading (that is, containing the minimal number of the derivatives)
gravitational correction to the quark condensate, Eq.~(\ref{le}). Finally, using 
the gravitational metric of a constant-energy-density domain, we 
evaluated the correction to the chiral condensate both inside and outside the domain [cf.
Eqs.~(\ref{in}) and (\ref{out})].
Outside the domain, this correction is extremely suppressed due to the smallness 
of the gravitational constant $G$ [cf. Eq.~(\ref{out1})], whereas inside the domain the correction could only 
get relevant at the distances to the center of the domain as small as those given by Eq.~(\ref{in1}). 
Such distances are, however, 20 orders of magnitude smaller than the macroscopically large radius of the domain, 
which makes the obtained correction numerically irrelevant.
The smallness of the correction, that the chiral condensate receives in the 
gravitational field of a domain, completely validates our use of the flat-space chiral condensate.

In conclusion, we put forward a conjecture that the domains of coherently excited pions could have been 
created in the early Universe. Such domains are hot
(as the estimates made in this paper remain valid up to temperatures of order 25~MeV) and stable against the gravitational collapse up to the 
maximum radius of about 14~km. Moreover, since the decay width of the coherent pionic states into photons
is negligibly small [cf. Eq.~(\ref{wd})], these domains cannot evaporate by means of the
electromagnetic radiation, which leads to their thermal insulation from the outer world. 
Since one can also argue for the stability of the 
coherent pionic states against the strong and weak decays~\cite{MoreRep,rep}, such encapsulated domains can have 
had a chance to survive till the present time, remaining however dark to external observers.

\begin{acknowledgments}
We thank N.~O.~Agasian for the useful correspondence.
The work of D.A. was supported by the Portuguese Foundation for Science and Technology
(FCT, program Ci\^encia-2008) and by 
the Center for Physics of Fundamental Interactions (CFIF) at Instituto Superior
T\'ecnico (IST), Lisbon. A.N. would like to thank the CFIF members for the warm hospitality extended to him during his stay in Lisbon. The work of A.N. was supported by the
State Corporation of Russian Federation ``Rosatom'', by the FCT
(grant PTDC/FIS/70843/2006-Fi\-si\-ca), by the German Research Foundation (DFG, grant 436 RUS 113/991/0-1), by the Russian Foundation for Basic Research (RFFI, grants
RFFI-09-02-91342-NNIOa and RFFI-09-02-00629a), as well as by the nonprofit Dynasty foundation and ICFPM.
\end{acknowledgments}

\appendix

\section{A world-line representation of the quartic self-interaction of matter fields}

In this Appendix, we recall the known world-line representation of the Abelian Higgs model with a non-local quartic interaction. Such a model is a bosonic prototype of the GNJL model. The part of its action containing the 
complex-valued Higgs field $\Phi(x)$ reads 
\be
S=\int_x\left(|D_\mu\Phi|^2+m^2|\Phi|^2\right)+\int_{x,y}|\Phi(x)|^2 K(x-y) |\Phi(y)|^2,
\label{A1}
\ee
where $D_\mu=\partial_\mu-igA_\mu$ is the covariant derivative.
To integrate over the Higgs field, one applies 
to the quartic term the Hubbard--Stratonovich transformation. Since it is the absolute value $|\Phi|$,
which enters the quartic term, the auxiliary scalar field $\sigma(x)$ is real-valued:
\be
{\rm e}^{-\int_{x,y}|\Phi(x)|^2 K(x-y) |\Phi(y)|^2}=\int 
{\cal D}\sigma{\,}{\rm e}^{-\int_x\left(\frac14\sigma K^{-1}\sigma+i\sigma|\Phi|^2\right)},
\label{A2}
\ee
where $K^{-1}$ is a local operator, whose Green function is $K$, that is $K^{-1}(x)K(x)=\delta(x)$.
(In particular, for an ordinary Abelian Higgs model with the local quartic interaction $K(x)=\kappa\delta(x)$,
the operator $K^{-1}$ is just a number: $K^{-1}=1/\kappa$.) Integration over the Higgs field then yields
\be
\int {\cal D}\Phi{\cal D}\Phi^{*}{\rm e}^{-S}=\int {\cal D}\sigma{\,}{\rm e}^{-\frac14\int_x
\sigma K^{-1}\sigma-\Gamma[A_\mu,\sigma]}\simeq\exp\left\{-\int {\cal D}\sigma{\,}{\rm e}^{-\frac14\int_x
\sigma K^{-1}\sigma}{\,}\Gamma[A_\mu,\sigma]\right\},
\label{A3}
\ee
where $\Gamma[A_\mu,\sigma]={\rm tr}{\,}\ln(-D_\mu^2+m^2+i\sigma)$ is the effective action.
The last approximation in Eq.~(\ref{A3}) was made analogously to the one-loop approximation for the effective action
with respect to the gauge field ($A_\mu$ in the present example), adopted in the bulk of the paper.
It means that the effective action 
describes a loop of the Higgs field with an arbitrary number of external lines of the gauge field, but 
does not describe gauge-field exchanges inside the loop and/or 
gauge-field interactions of two and more such loops. 
Similarly, Eq.~(\ref{A3}) assumes the lowest-loop approximation for the 
effective action with respect to the $\sigma$-field. We consider now 
the averaged effective action 
\be
\Gamma[A_\mu]\equiv\int {\cal D}\sigma{\,}{\rm e}^{-\frac14\int_x
\sigma K^{-1}\sigma}{\,}\Gamma[A_\mu,\sigma]
\label{A4}
\ee
and apply the world-line representation to write [cf. Eq.~(\ref{effectivea2})]
\be
\Gamma[A_\mu,\sigma]=2V\int_0^\infty\frac{ds}{s}{\rm e}^{-m^2s}\int_P {\cal D}z_\mu
{\rm e}^{-\int_0^s d\tau\left[\frac14 \dot z_\mu^2+i\sigma(z(\tau))\right]}{\,}W[z_\mu, A_\mu].
\label{A5}
\ee
Here, $W[z_\mu, A_\mu]=\exp\left(ig\int_0^s d\tau A_\mu\dot z_\mu\right)$ is the Abelian phase factor 
along the loop, and the overall factor of 2 is because of 
the two (real and imaginary) components of the Higgs field. Integrating now over the $\sigma$-field, one obtains
the following effective action:
\be
\Gamma[A_\mu]=2V\int_0^\infty\frac{ds}{s}{\rm e}^{-m^2s}\int_P {\cal D}z_\mu
{\rm e}^{-\frac14\int_0^s d\tau\dot z_\mu^2-\int_0^s d\tau\int_0^s d\tau' K[z(\tau)-z(\tau')]}{\,}
W[z_\mu, A_\mu].
\label{A6}
\ee
As one can see, because of the two-point interaction produced by the nonlocal kernel $K(x)$, we have arrived at 
the two-loop effective action (instead of the one-loop effective action, which one gets in the same lowest-loop 
approximation
when accounting only for the interaction of the Higgs field with the gauge field), which 
cannot be calculated analytically. In the particular case of a short-range (or a local) kernel, the term 
$\int_0^s d\tau\int_0^s d\tau' K[z(\tau)-z(\tau')]$ suppresses in the world-line integral self-intersecting 
trajectories. For this reason, the Abelian Higgs model with a {\it short-range} 
nonlocal quartic interaction is known to be a field-theoretical 
representation for the ensemble of {\it self-avoiding} bosonic random walks~\cite{zj}.

\section{The gravitational field of a spherically-symmetric object with a constant energy density}

In this Appendix, we summarize some known facts~\cite{w}
about the gravitational field, the energy density $\varepsilon$, 
and the pressure $p$ of spherically-symmetric objects, with a particular emphasis on those with the constant energy density. Throughout this Appendix, for terminological simplicity, we will call such objects stars.
Usually, one assumes that the matter forming a star is a perfect fluid. This implies 
the energy-momentum tensor of the form
$T^{\mu\nu}=(p+\varepsilon)u^\mu u^\nu-pg^{\mu\nu}$,
with $u^\mu(x)$ being the four-velocity of the fluid, such that $g_{\mu\nu}u^\mu u^\nu=1$.
In the local rest frame of the 
fluid, where $u_\mu=(\sqrt{g_{00}},{\bf 0})$, the energy-momentum tensor is diagonal: 
$$
T^\mu{\,}_\nu=(p+\varepsilon)u^\mu u_\nu-p\delta^\mu{\,}_\nu={\,}{\rm diag}{\,}(\varepsilon,-p,-p,-p).
$$ 
Furthermore, since the star is spherically 
symmetric, the $x$-dependence of $\varepsilon$ and $p$ is reduced to their dependence on the 
spatial distance to the center, $r$. For this reason, one uses the 3D spherical coordinates $(r,\theta,\phi)$
to find the metric components $g_{00}={\rm e}^{a(r)}$ and $g_{11}=-{\rm e}^{b(r)}$, while the two other components read
$g_{22}=-r^2$ and $g_{33}=-r^2\sin^2\theta$. Outside the star, that is for $r>R$, where $R$ is the star radius, the
energy-momentum tensor $T^{\mu\nu}$ vanishes, and one 
obtains the Schwarzschild metric
\be
a(r)=-b(r)=\ln \left(1-\frac{r_g}{r}\right)
\label{B1}
\ee
with the  
Schwarzschild radius $r_g\equiv 2GM$, where $G$ is the gravitational constant, and 
$M=4\pi\int_0^R dr r^2\varepsilon(r)$ is the full energy of the star.

For $r<R$, it is convenient to introduce the energy inside the radius $r$, given by ${\cal M}(r)=4\pi 
\int_0^r dr' r'^2\varepsilon(r')$. The Einstein equation ${\cal R}^0{\,}_0-\frac12{\cal R}=8\pi GT^0{\,}_0$
reads 
$$
\frac{{\rm e}^{-b}}{r}\left(\frac{db}{dr}-\frac1r\right)+\frac{1}{r^2}=8\pi G\varepsilon.
$$
Its solution $b(r)=-\ln (1-\frac{2G{\cal M}}{r})$ clearly goes over to the function~(\ref{B1}) at $r=R$. 
The function $a(r)$ inside the star can be found from the covariant conservation of the energy-momentum 
tensor, $\nabla_\mu T^{\mu\nu}=0$, which explicitly reads
\be
-\partial_\mu p\cdot g^{\mu\nu}+\partial_\mu\left[(p+\varepsilon)u^\mu u^\nu\right]+
(p+\varepsilon)\left(\Gamma_{\lambda\mu}^\mu u^\lambda u^\nu+\Gamma_{\lambda\mu}^\nu 
u^\mu u^\lambda\right)=0.
\label{B2}
\ee
One assumes furthermore the hydrostatic-equilibrium condition, which implies the $x_0$-independence of 
not only $p$ and $\varepsilon$ but also of $u_\mu=(\sqrt{g_{00}},{\bf 0})$. In particular, this condition yields
$\partial_\mu[(p+\varepsilon)u^\mu u^\nu]=\partial_0[(p+\varepsilon)u^0 u^0]=0$. Multiplying the rest 
of Eq.~(\ref{B2}) by $g_{\rho\nu}$, one has
$$
\frac{\partial_\rho p}{p+\varepsilon}=\Gamma_{\lambda\mu}^\mu u^\lambda u_\rho+
g_{\rho\nu}\Gamma_{\lambda\mu}^\nu u^\mu u^\lambda.
$$
The first term on the right-hand side of this equation vanishes, while the second term 
leads to the desired equation
\be
\frac{d\ln g_{00}}{dr}=-2\cdot\frac{dp/dr}{p+\varepsilon}.
\label{B3}
\ee
The solution to this equation,
\be
g_{00}(r)=g_{00}(R)\cdot\exp\left[2\int_r^R dr'{\,}\frac{dp/dr'}{p+\varepsilon}\right]~~~ 
{\rm with}~~~  g_{00}(R)=1-\frac{r_g}{R}
\label{B4}
\ee
defines the temporal component of the metric inside the star, as well as the function $a(r)=\ln g_{00}(r)$.
Similar to the function $b(r)$, such a function $a(r)$ goes over to Eq.~(\ref{B1}) at $r=R$.

The equation for $p(r)$ can be obtained from the Einstein equation 
${\cal R}^1{\,}_1-\frac12{\cal R}=8\pi GT^1{\,}_1$, which is equivalent to 
$$
\frac{da}{dr}={\rm e}^b\left(\frac1r+8\pi Grp\right)-\frac1r.
$$
To this end, it suffices to use $da/dr$ given by Eq.~(\ref{B3}), $\frac{da}{dr}=-2\frac{dp/dr}{p+\varepsilon}$,
and the relation ${\rm e}^b=(1-2G{\cal M}/r)^{-1}$. That yields the so-called Tolman--Oppenheimer--Volkoff 
equation for $p(r)$~\cite{w}:
\be
-\frac{dp}{dr}=\frac{G\varepsilon{\cal M}}{r^2}\left(1-\frac{2G{\cal M}}{r}\right)^{-1}\left(1+
\frac{p}{\varepsilon}\right)\left(1+\frac{4\pi r^3p}{\cal M}\right).
\label{B5}
\ee
Together with the equation $\frac{d{\cal M}}{dr}=4\pi r^2\varepsilon$ 
and the equation of state, it forms a set of 
three equations for the three unknown functions, that are $p$, $\varepsilon$, and ${\cal M}$.
Furthermore, plugging Eq.~(\ref{B5}) into Eq.~(\ref{B4}), 
one also obtains the temporal component of the metric inside 
the star in terms of the functions $p$ and ${\cal M}$:
\be
g_{00}(r)=g_{00}(R)\exp\left[-2G\int_r^R\frac{dr'}{r'^2}\left(1-\frac{2G{\cal M}}{r'}\right)^{-1}
\left({\cal M}+4\pi r'^3p\right)\right].
\label{B6}
\ee
In the particular case $\varepsilon={\,}{\rm const}$, 
one has ${\cal M}=\frac{4\pi}{3}\varepsilon r^3$, and 
Eq.~(\ref{B5}) can be integrated analytically to yield 
\be
p(r)=\varepsilon\cdot\frac{\sqrt{1-\frac{8\pi}{3}\varepsilon Gr^2}-
\sqrt{1-\frac{8\pi}{3}\varepsilon GR^2}}{3\sqrt{1-\frac{8\pi}{3}\varepsilon GR^2}-
\sqrt{1-\frac{8\pi}{3}\varepsilon Gr^2}},
\label{B7}
\ee
where the boundary condition $p(R)=0$ has been taken into account. 
For the denominator in this formula not to vanish 
at any $r<R$, one imposes the requirement $3\sqrt{1-\frac{8\pi}{3}\varepsilon GR^2}-1>0$, which defines the 
upper limit for the star radius: 
\be
R\le \frac{1}{\sqrt{3\pi\varepsilon G}}.
\label{B8}
\ee
Finally, according to Eq.~(\ref{B6}), the temporal component of the metric inside the star of 
a constant energy density reads
\be
g_{00}(r)=\left(1-\frac{r_g}{R}\right)\exp\left[-8\pi G\int_r^R\frac{dr'r'}{1-
\frac{8\pi\varepsilon G}{3}r'^2}\left(\frac{\varepsilon}{3}+p(r')\right)\right],
\label{B9}
\ee
where the function $p(r')$ is given by Eq.~(\ref{B7}).


\begin{thebibliography}{99}

\bibitem{ds} H.~G.~Dosch and Yu.~A.~Simonov, Phys.\ Lett.\  B {\bf 205}, 339 (1988).

\bibitem{svz} M.~A.~Shifman, A.~I.~Vainshtein and V.~I.~Zakharov, Nucl.\ Phys.\  B {\bf 147}, 385 (1979).

\bibitem{AntRib} D.~Antonov and J.~E.~F.~T.~Ribeiro, Phys.\ Rev.\  D {\bf 81}, 054027 (2010).

\bibitem{NJL} Y.~Nambu and G.~Jona-Lasinio, Phys.\ Rev.\  {\bf 122}, 345 (1961).

\bibitem{results} 
S.~L.~Adler and A.~C.~Davis, Nucl.\ Phys.\ B {\bf 244}, 469 (1984); A.~Le Yaouanc, L.~Oliver, S.~Ono, O.~Pene and J.~C.~Raynal, Phys.\ Rev.\  D {\bf 31}, 137 (1985);   
C.~D.~Roberts and A.~G.~Williams, Prog.\ Part.\ Nucl.\ Phys.\ {\bf 33}, 477 (1994);
P.~Bicudo, J.~E.~F.~T.~Ribeiro, and J.~Rodrigues, Phys.\ Rev.\ C {\bf 52}, 2144 (1995); 
R.~Horvat, D.~Kekez, D.~Palle, and D.~Klabucar, Z.\ Phys.\ C {\bf 68}, 303 (1995); 
N.~Brambilla and A.~Vairo, Phys.\ Lett.\ B {\bf 407}, 167 (1997);
P.~Maris and C.~D.~Roberts, Phys.\ Rev.\ C {\bf 56}, 3369 (1997);
P.~Maris, C.~D.~Roberts, and P.~C.~Tandy, Phys.\ Lett.\ B {\bf 420}, 267 (1998);
P.~Maris and P.~C.~Tandy, Phys.\ Rev.\ C {\bf 60}, 055214 (1999); {\it ibid.}, {\bf 61}, 045202 (2000);
F.~J.~Llanes-Estrada and S.~R.~Cotanch, Phys.\ Rev.\ Lett. {\bf 84}, 1102 (2000); 
P.~Bicudo, Phys.\ Rev.\  C {\bf 67}, 035201 (2003);
N.~Ligterink and E.~S.~Swanson, Phys.\ Rev.\ C {\bf 69}, 025204 (2004).

\bibitem{fr} A.~Le Yaouanc, L.~Oliver, O.~Pene and J.~C.~Raynal, Phys.\ Rev.\  D {\bf 29}, 1233 (1984);
P.~Bicudo and J.~E.~F.~T.~Ribeiro, Phys.\ Rev.\  D {\bf 42}, 1611, 1625, 1635 (1990).

\bibitem{p} P.~Bicudo, S.~Cotanch, F.~J.~Llanes-Estrada, P.~Maris, J.~E.~F.~T.~Ribeiro and A.~Szczepaniak,
Phys.\ Rev.\  D {\bf 65}, 076008 (2002).

\bibitem{GMOR} M.~Gell-Mann, R.~J.~Oakes and B.~Renner, Phys.\ Rev.\  {\bf 175}, 2195 (1968).

\bibitem{GT} M.~L.~Goldberger and S.~B.~Treiman, Phys.\ Rev.\  {\bf 111}, 354 (1958).

\bibitem{Weinberg} S.~Weinberg, Phys.\ Rev.\ Lett.\  {\bf 17}, 616 (1966).

\bibitem{Adler} S.~L.~Adler, Phys.\ Rev.\  {\bf 137}, B1022 (1965).

\bibitem{w} For reviews, see: S.~Weinberg, ``Gravitation and cosmology'', Wiley \& Sons, 1972; K.~Yagi, T.~Hatsuda and Y.~Miake, ``Quark-gluon plasma'', Cambridge University Press, 2005.

\bibitem{MoreRep} P.~Bicudo, J.~E.~F.~T.~Ribeiro and A.~V.~Nefediev, Phys.\ Rev.\  D {\bf 65}, 085026 (2002);
P.~Bicudo and A.~V.~Nefediev, Phys.\ Rev.\  D {\bf 68}, 065021 (2003);
A.~V.~Nefediev and J.~E.~F.~T.~Ribeiro, Phys.\ Rev.\  D {\bf 70}, 094020 (2004).

\bibitem{rep} A.~V.~Nefediev and J.~E.~F.~T.~Ribeiro, Phys.\ Rev.\  D {\bf 67}, 034028 (2003).

\bibitem{replica2d} P.~Bicudo and A.~V.~Nefediev, Phys.\ Lett.\  B {\bf 573}, 131 (2003).

\bibitem{OH} See, for example: A.~A.~Osipov and B.~Hiller, Phys.\ Lett.\ B {\bf 539}, 76 (2002);
F.~J.~Llanes-Estrada, T.~Van~Cauteren and A.~P.~Martin, Eur.\ Phys.\ J.\ C {\bf 51}, 945 (2007).

\bibitem{ne} P.~Bicudo, N.~Brambilla, J.~E.~F.~T.~Ribeiro and A.~Vairo, Phys.\ Lett.\  B {\bf 442}, 349 (1998).

\bibitem{WL} M.~G.~Schmidt and C.~Schubert, Phys.\ Lett.\  B {\bf 318}, 438 (1993); Phys.\ Lett.\  B {\bf 331}, 69 (1994); for reviews, see: M.~Reuter, M.~G.~Schmidt and C.~Schubert, Annals Phys.\  {\bf 259}, 313 (1997);
C.~Schubert, Phys.\ Rept.\  {\bf 355}, 73 (2001).

\bibitem{wq} M.~D'Elia, A.~Di Giacomo and E.~Meggiolaro, Phys.\ Lett.\  B {\bf 408}, 315 (1997); 
G.~S.~Bali, N.~Brambilla and A.~Vairo, Phys.\ Lett.\  B {\bf 421}, 265 (1998).

\bibitem{en} E.~Meggiolaro, Phys.\ Lett.\  B {\bf 451}, 414 (1999).

\bibitem{smalla} For a recent discussion, see: B.~Z.~Kopeliovich, I.~K.~Potashnikova, B.~Povh and I.~Schmidt,
Phys.\ Rev.\  D {\bf 76}, 094020 (2007);
A.~M.~Badalian, A.~V.~Nefediev and Yu.~A.~Simonov,  JETP Lett.\  {\bf 88}, 558 (2008); 
Phys.\ Rev.\  D {\bf 78}, 114020 (2008).


\bibitem{sb} See, for example: D.~Antonov, L.~Del Debbio and A.~Di Giacomo, JHEP {\bf 08}, 011 (2003); 
D.~Antonov and A.~Di Giacomo, JHEP {\bf 03}, 017 (2005).

\bibitem{1} For a review, see: N.~O.~Agasian, Phys.\ Atom.\ Nucl.\  {\bf 68}, 723 (2005).

\bibitem{gl} N.~K.~Glendenning, ``Compact stars'', 2nd edn., Springer-Verlag, 2000.

\bibitem{gr} I.~T.~Drummond and S.~J.~Hathrell, Phys.\ Rev.\  D {\bf 22}, 343 (1980);
F.~Bastianelli, J.~M.~D\'avila and C.~Schubert, JHEP {\bf 03}, 086 (2009).

\bibitem{ll} L.~D.~Landau and E.~M.~Lifshitz, ``Classical theory of fields'', 6th edn., Pergamon Press, 1988.

\bibitem{zj} See, for example: J.~Zinn-Justin, ``Quantum field theory and critical phenomena'', 4th edn., Clarendon Press, 2002.

\end{thebibliography}
\end{document}